# Classical analysis of the rotational dynamic of spiral galaxies: Quo Vadis Dark Matter?

E. López-Sandoval
Centro Brasileiro de Pesquisas Físicas,
Rua Dr. Xavier Sigaud, 150
CEP 22290-180, Rio de Janeiro, RJ, Brazil.
E-mail: sandoval@cbpf.br

In this paper we study like a "stellar dynamic" model, the stars' rotational dynamics of spiral galaxies: with a distribution of its own mass, rotating around its centre (with a higher density), by means of a classical calculus we obtain the rotational velocities of a star around its rotation axis (center of the galaxy), considering that it is inside a smoothed distribution of matter. Two kinds of matter distribution are supposed: one with constant density, and other with radial and angular distribution. The stars are supposed to be particles and their distribution in the galaxy is modeled as a matter distribution inversely proportional to its distance from its centre. Two types of galaxy symmetry are also considered: spherical and oblate ellipsoidal. Using only classical mechanics arguments it is shown that the calculated velocity distribution inside a galaxy is similar to that obtained from astronomical observations, without the necessity of suppose the existence of dark matter or other phenomena.

## 1 Introduction

*"It would not have to forget that all this speech about the nature of the universe implies a tremendous extrapolation, which is a very dangerous operation".*

*W. de Sitter (1931).*

In some studies carried out by Rubin et al. [1,2] it was found that the rotational speed of spiral galaxies of Hubble type (Sa, Sb and Sc) grew fast in distances near to its centre. But for distances faraway of the centre of the galaxies and in their periphery, the velocity of the stars grows slowly and tends to a maximum speed limit, which is higher than the velocities for lower distances. Their results also showed that this velocity distribution does not depend on the Hubble type of spiral galaxies. Moreover, the values for the velocities distributions were greater for those of type Sa than for types Sb and Sc, being Sa less extended in its equatorial radius with regard to the Sb and Sc, and having the same luminous intensity. The same conditions are satisfied for the velocities of Sb with respect to those of Sc. For the galaxies of the same type, they showed that those of greater brightness had greater rotation velocities [2].

Explanations for the high rotation velocities of the stars of spiral galaxies in their periphery have been of two types: a) the presence of dark matter, whose properties at the present are not very well known, but that was necessary to propose its existence because it permits to explain why the stars are not ejected out of the galaxies due that the total mass of these are not sufficient to retain them rotating around their centre, and b) the Newton laws no longer comply to this scale, and it is necessary to make a correction.

The first proposal was carried out by Rubin et al. [1], and provoked great experimental and theoretical efforts in the search of the properties and structure of this exotic matter. According to this hypothesis the galaxy are covered with a substantial amount of matter far from the centre of the galaxy, and is known like dark matter halo. This form of matter does not emit or reflect electromagnetic radiation. The quantity and structure of this form of matter have been widely studied [3-6]. Although a rotund experimental test about their existence has not been found.

The second explanation, stated by Milgron [7] proposes a modification of the second Newton's law for a weak gravity. In the same way, Disney [8] suggests modifications to the Newton's law of Universal Gravitation.

The Milgrom modification concern write the acceleration of the body proportional to a new one universal constant, and it reads as:

$$F = m\mu\left(\frac{a}{a_0}\right)a, \quad (1)$$

Where $a_0 = 1.2x10^{-10}$ ms$^{-2}$, and $\mu(x) = 1$ if $x >> 1$, or $\mu(x) = x$ if $x << 1$. If $a >> a_0$ for all physical effects then $\mu(a/a_0) = 1$, and from (1) we have $F = ma$, as it usually known. This theory explain that the acceleration of a particle, at low values of acceleration, is not linearly proportional to the force because, according to him, the inertial mass of a body depends on the field strength which it is subjected [7].

The proposal of Disney is the modification of the inverse-square law:

$$F = Gm_1m_2\left(\frac{1}{d^2}\right), \quad (2)$$

and the modification suggested consists in adding an extra term to the inverse of the distance, *i.e.*:

$$F = Gm_1m_2\left(\frac{1}{dD_0} + \frac{1}{d^2}\right), \quad (3)$$

where $D_0$ is some very large fixed distance [8] .In this way, for $d << D_0$, the extra term is insignificant, and the Eq. (3) recovers the already known inverse-square law for the Newtonian gravitation. But, when $d \approx D_0$ or $d > D_0$, the extra term begins to be significant and the intensity of the interaction force among the bodies will increase. This increment in the force would be able to explain the cohesion of the periphery stars to the galaxy in spite of their high rotational speed. Although the Disney proposal is interesting, it modifies the inverse-square law that is the form as the gravitational and electric fields behave in space. Some theoretical analysis have shown that the inverse-square law provides the stability of planetary orbits and atomic ground states, the propagation of the electromagnetic waves without reverberation occurs, the fundamental constants of nature, and all these properties are the result of a three dimensional space [9].

In this paper we make an idealized analysis for the rotational dynamic of stars, in a distribution of its own mass, that are rotating around its center with a higher density, like happen in the spiral galaxies.

## 2. The model

To carry out the theoretical calculation of the rotation velocity of the stars in a galaxy, generally it is applied the Universal Gravitation law´s (or Kepler´s third law), where the force on the stars decreases with the inverse of the square distance of the rotation axis, just as it is used for calculate the rotation of the planets around the Sun. In the case of the solar system, this procedure is correct, because the greater mass of the solar system is concentrated on the sun (*99.8* per cent), and although there is interaction among the planets, their effect is small with respect to the interaction with the sun. In calculations for the dynamic of the solar system, the interactions between the planets are taken as tiny perturbations. But for a system as large as a galaxy, which is composed on the order of $10^{11}$ stars faraway among them by distances of light-years, the gravitational force among stars is very weak, and although the stars also rotate around a rotational axis located in the galaxy centre, the mass is not all concentrated on this, but it is distributed by the entire galaxy. Nevertheless, we known that this distribution of matter is more concentrated in its centre, with a higher density with respect another part of the galaxy[*], and diminishing radially as we moved away of this one, and we approached to the spiral arms.

Misunderstand arises, it's supposed, by considering that gravitational interaction between the stars in a galaxy is very weak and therefore it has little influence in its dynamic behavior. In addition, by a reductionistic vision that carried to an erroneous interpretation of the Gauss's law, where is considered that the whole mass of the galaxy is concentrated in one point in its centre, instead of considering the internal gravitational field of the galaxy as that of system of particles with a inhomogeneous distribution of matter.

For example, if we consider the internal gravitational field of a system of particles with a homogeneous or spherical radial distribution of matter, and in agreement to the classical calculation with spherical shells that is carried out, the effect of the force of the exterior mass to its internal radio is annulled by symmetry, as like this external mass would not exist, or as if was cut off, and only receiving influence of the force exercised by the internal mass concentrated on a point in its centre. In fact, the effect of the interaction of the exterior mass is not annulled, but its net average effect is zero due to its symmetry, and only remains the force of the remaining internal mass. But this scheme is valid only for a spherical symmetry and for each particle the net force is different depending of its internal distribution of mass locked up by a Gaussian surface. For other symmetries as those of the galaxies, the internal force of the entire distribution is not annulled and acting over each particle.

[*] It have been detected in the centre of many galaxies, super massive black holes with a concentration of mass until $10^{8}$ suns [10], reaching until *0.1* per cent of the mass of the entire galaxy.

We will model this distribution like a function that decreases inversely proportional to the *n*th power of the distance from its centre, with $0< n\leq 1.8$. We chose this kind of distribution because it has been discovered, by means of correlation functions, that the distribution of galaxies in the large scale structure of the universe follows this behavior with $n=1.8$ [11,12]. It's supposed that the galaxies are cluster with this type of structure because the gravity could induce such a power law behavior [12]. We will present like hypothesis that stars in galaxies follow a similar behavior. Besides, we considered thus because this has a singular point that describes the influence of the black hole on its rotational velocities of the stars.

By the great number of stars in galaxies and the great distance between them (order of light-years), and for the purpose of this work, we will consider an ideal situation and neglect the physical size of star and to consider particles that interact gravitationally between then. Each particle (where each one represents a star with a similar mass to the sun) contributes to the overall gravitational field. We only will consider their interaction as internal forces in the entire galaxy, i.e. we will not consider external forces due to other galaxies for this approximation. Because there are many stars in a typical spiral galaxy ($10^{11}$) and we don't need to know the precise location of each star for to calculate its net force, we can to do an approximation. For to obtain an excellent estimation is necessary only to replace this distribution of individual stars by a smoothed continuum function of density. With this, we neglect local gravitational effects, and each particle follow a ´collisionless dynamic' under a smoothed gravitational field, in a stationary system. In contrast with electric force with charged particles, gravitational force isn't annulled, and therefore it is a cumulative force.

Therefore, for to obtain this global and average net force over each star we need to use the Gauss´ law.

We will make an analysis using the Gauss's law for to obtain the potentials of a distribution of matter with spherical symmetry, and next applying a quadripotential approximation (QA) to these [13,14], for to obtain the potentials of a distribution with ellipsoidal symmetry that will take into account an approximated geometry of the spiral galaxies. The gravitational field will be calculated from this distribution, if first calculate its potential generated by its mass distribution. This gravitational field has a distribution by all the system, and should be taken into account for the calculation of the internal dynamic of each star in the galaxy. Our study will be an idealization, but that we will serve us to obtain an average calculation of the velocities of rotation, helping us to test our hypothesis.

## 3. Classical Analysis

We consider the dynamic evolution of two types of smoothed mass distributions: homogeneous distribution and with a density that decreases inversely proportional to the *n*th power of the distance from its centre (*n=0.5, 1, 1.5*, and *1.8*). This means that a greater power increased the density in its center, and decrease out of this. Also with two types of symmetries: spherical and oblate ellipsoidal. In this framework, we have to calculate the distribution of the force or gravitational field of these two distributions, applying the gradient to its respective potentials.

### 3.1 Spherical distribution of matter

In the first case, for a homogenous spherical distribution of matter with a constant density,

$$\rho = \frac{M}{4\pi R^3/3},$$

where *M is* the mass of the galaxy, that we approximated to $10^{11}$ times the mass of the sun, that is the average number of stars by galaxy, i.e., $1.99 \times 10^{41}$ kg; and *R* is the average radius. We can directly apply the Gaussian law for to obtain the gravitational field:

$$\oint g \cdot dA = -4\pi G \iiint \rho dV \,. \tag{4}$$

From this one, we can obtain the gravity force for any particle inside the distribution:

$$g = -\frac{GM}{R^3} r\hat{r} \,, \tag{5}$$

making a line integral in the gravitational field, we obtain the internal potential:

$$\Phi = -\frac{GM}{2R^3}\left(3R^2 - r^2\right). \tag{6}$$

We know that the time of rotation of our sun around Milky Way galaxy is hundreds of millions of years (approximately 2 x $10^8$ years), then for the time of human life, we can consider that the dynamics is not modified significantly at the moment of our observation. Consequently, we can consider that $r$ and $\theta$ do not vary significantly and there are not angular or translational accelerations. Therefore, in this interval of time we can carry out the approximation that the centrifugal force is in equilibrium with the component of the gravity in the same direction. According to the Newton's second law,

$$g = a_c \,. \tag{7}$$

These stars rotate around its axis, and the centrifugal acceleration can be written as:

$$a_c = \frac{V^2}{r}\hat{r} \quad . \tag{8}$$

Using the equation (7) we obtain

$$\frac{V^2}{r}\hat{r} = \frac{GMr}{R^3}\hat{r}, \tag{9}$$

and therefore:

$$V = \left(\frac{GM}{R}\right)^{1/2}\frac{r}{R}, \tag{10}$$

where $G$ is the gravitational constant, $M$ the mass of the galaxy, that we approximated to $10^{11}$ times the mass of the sun, that is the average number of stars by galaxy, *i.e.*, 1.99 x $10^{41}$ kg; and $R$ is the average radius of 4.8274 kpc. We took this radius from the average value of an oblate ellipsoid whose dimensions are $a = 0.5$ kpc, $b = 15$ kpc and $c = 15$ kpc. This is the measures with which we are going to approximate the galaxy with the volume of an ellipsoid, in the next part. We observed from Eq. (10) that the rotational velocity for this symmetry and corresponding distribution, grows linearly with the radial distance from its center, reaching approximately 300 km/s near its periphery. The velocity distribution in the radial direction is shown in figure 1.

Now, we suppose that the density is not constant and that it grows how the inverse of the radius elevated to some $n$-power. We begin with the power $n = 1/2$:

$$\rho = \left(\frac{5M}{8\pi R^{5/2}}\right)\frac{1}{r^{1/2}}. \tag{11}$$

Applying the Gauss's law for this radial spherical distribution, we obtain for the gravitational force:

$$g = -\frac{GM}{R^{5/2}} r^{1/2}\hat{r}, \tag{12}$$

and troughout a line integral, we obtain its potential

$$\phi = -\frac{GM}{3R^{5/2}}\left(5R^{3/2} - 2r^{3/2}\right). \tag{13}$$

As in the previous case, we use the Newton's second law (7) to analyze the rotational dynamics with the field of gravitation $g$ of Eq.(12):

$$\frac{V^2}{r} = \frac{GM}{R^{5/2}} r^{1/2}, \tag{14}$$

and so that the velocity distribution results:

$$V = \left(\frac{GM}{R}\right)^{1/2} \left(\frac{r}{R}\right)^{3/4}. \tag{15}$$

We generate a series of values that allow us show the graph of the distribution of matter and velocities in the radial direction. We calculate these quantities at regular intervals from the origin, at each 0.1 kpc, until to completing the length of 4.8 kpc. The density and velocities distribution are shown by the curves 2(a) and 3(a) respectively.

In a similar way, we can calculate the field of velocities for other distributions in whose center its density grows still faster: for $n = 1$,

$$\rho = \left(\frac{M}{2\pi R^2}\right) \frac{1}{r}. \tag{16}$$

This distribution of matter has an internal gravitational field:

$$g = -\frac{GM}{R^2} \hat{r}, \tag{17}$$

and the potential reads:

$$\phi = -\frac{GM}{R^2} (2R - r). \tag{18}$$

Now, we calculate the velocity of rotation inside this distribution using (7):

$$\frac{V^2}{r} = \frac{GM}{R^2}, \tag{19}$$

and finally the  velocity field results:

$$V = \left(\frac{GM}{R}\right)^{\frac{1}{2}} \left(\frac{r}{R}\right)^{1/2}. \tag{20}$$

The corresponding graphs of the density (Eq. (16)) and velocity distributions (Eq. (20)) are given by the curves 2(b) and 3(b) shown in the figures 2 and 3 respectively.

In one more step in our study, we present the calculations for a distribution still more concentrated in its center, $n = 3/2$:

$$\rho = \left(\frac{3M}{8\pi R^{3/2}}\right) \frac{1}{r^{3/2}}. \tag{21}$$

The density graph is showed in 2(c). The internal gravitational field is:

$$g = -\frac{GM}{R^{3/2}} \frac{1}{r^{1/2}} \hat{r}, \tag{22}$$

and the potential is writes like:

$$\phi = -\frac{GM}{R^{3/2}} \left(3R^{1/2} - 2r^{1/2}\right). \tag{23}$$

The velocity distribution, from the Newton's law, reads

$$V = \left(\frac{GM}{R}\right)^{\frac{1}{2}} \left(\frac{r}{R}\right)^{1/4}. \tag{24}$$

Its curve is shown in the figure 3(c).

Finally, we studied a distribution that goes inversely radius inversely proportional to $n = 9/5 \approx 1.8$[†],

$$\rho = \left( \frac{6M}{20\pi R^{6/5}} \right) \frac{1}{r^{9/5}}, \tag{25}$$

whose curve is showed in 2(d). Its gravitational field is

$$g = -\left( \frac{GM}{R^{6/5}} \right) \frac{1}{r^{4/5}} \hat{r}, \tag{26}$$

and its potential reads:

$$\phi = -\frac{GM}{R^{6/5}} \left( 6R^{1/5} - 5r^{1/5} \right). \tag{27}$$

Now as in the previous cases, we calculate its rotational velocity:

$$V = \left( \frac{GM}{R} \right)^{1/2} \left( \frac{r}{R} \right)^{1/10}. \tag{28}$$

This distribution of velocities represents the limit case of the speed of rotation for these distribution and symmetry. The curve 3(d) represents this velocity.

From the analysis of the results sketched in Figs. 1-3, we see that for a spherical distribution of mass, the speed of stars depends of its symmetry form and its density. For these densities with radial function, our calculations are more realistic than for a homogeneous distribution, but it continues being an idealization picture for the form that the spiral galaxies have.

---

[†] We chose this distribution because it has been discovered, by means of correlation functions, that the distribution of galaxies in the large scale structure of the universe follows this behavior in clusters and super cluster [11, 12]. It's supposed that the stars in galaxies follow a similar behavior.

### *3.2 Oblate ellipsoidal distribution of matter*

A more realistic approximation is to consider an ellipsoidal distribution of extreme oblate form. Thus, we could to approximate in an ideal manner to the disc form that these galaxies have. For this purpose, we will use the potentials of the spherical distribution of mass, for each one of the previously distributions of matter calculated making a QA for each potential.

We begin with the potential of the distribution with constant density, and we obtain the internal potential for an ellipsoidal homogenous distribution [13, 14]:

$$\Phi = -\frac{GM}{2R^3}\left(3R^2 - r^2\right) - \frac{GM}{R^3}\frac{\eta(2-\eta)r^2}{5}\left(1 - 3Cos^2\theta\right), \tag{29}$$

where $M = 4\pi R^3\rho/3 = 4\pi R_e^2 R_p \rho/3$. Although the size of the radius of the sphere is modified to two semi-axis of the equator and one of pole, the volume continues being the same one, with the same constant density, for an average radius $R$ of the sphere. The parameter $\eta$ is obtained from the relation $R_p = R_e\left(1-\eta\right)$, resulting in $\eta = 1 - R_p/R_e$. This value is under the condition $\eta < 1$, since we are taking the quadratic approximation due to that the proportion between radio of the equator and the pole of the galaxies comes to be 15 times or more. Applying the gradient in spherical coordinates to this potential, the internal gravitational vectorial field for this distribution of matter reads:

$$g = -\frac{GMr}{R^3}\left\{\left[1 - \frac{2}{5}\eta(2-\eta)(1 - 3Cos^2\theta)\right]\hat{r} - \frac{6}{5}\eta(2-\eta)Sin\theta Cos\theta\hat{\theta}\right\}, \tag{30}$$

where $M$ is the mass of galaxy, and $R$ the radius of the sphere, or the average radius of the ellipsoid. The vectors $\hat{r}$ and $\hat{\theta}$ are unitary vectors in spherical coordinates. This field determines the internal dynamics of the galaxy for any proportion among its axis (polar and equatorial) for this distribution of matter with a constant density.

Now we calculate the rotational velocity of stars around its polar axis on the equatorial plane. From Eq. (7), of the Newton's second law, we have

$$g \cdot \hat{\rho} = \frac{V^2}{\rho},$$

where $\hat{\rho} = (Sin\theta)\hat{r} + (Cos\theta)\hat{\theta}$ is a unitary polar vector in function of the unitary vectors in spherical coordinates. Then

$$\frac{GMr}{R^3}\left\{\left[1 - \frac{2}{5}\eta(2-\eta)(1 - 3Cos^2\theta)\right]\hat{r} - \frac{6}{5}\eta(2-\eta)Sin\theta Cos\theta\hat{\theta}\right\} \cdot \left[(Sin\theta)\hat{r} + (Cos\theta)\hat{\theta}\right] = \frac{V^2}{rSin\theta}$$

And therefore

$$\frac{GMr}{R^3}\left\{\left[1 - \frac{2}{5}\eta(2-\eta)(1 - 3Cos^2\theta)\right]Sin\theta - \frac{6}{5}\eta(2-\eta)Sin\theta Cos^2\theta\right\} = \frac{V^2}{rSin\theta} \qquad (31)$$

Clearing $V$, in function of the variables $r$ and $\theta$, reads:

$$V(r,\theta) = \left(\frac{GM}{R^3}\right)^{1/2} rSin\theta\left[1 - \frac{2}{5}\eta(2-\eta)\right]^{1/2}, \qquad (32)$$

or substituting in $R^3$ the semi-axis polar $R_p$, and the equatorial $R_e$, we have:

$$V(r,\theta) = \left(\frac{GM}{R_p}\right)^{1/2}\left(\frac{rSin\theta}{R_e}\right)\left[1 - \frac{2}{5}\eta(2-\eta)\right]^{1/2}. \qquad (33)$$

For to obtain the graphic of $V(r,\theta)$, we take the parameters $R_p = 0.5$ kpc, $R_e = 15$ kpc, and therefore $\eta(2-\eta) = 0.966$, that is the dimensions of a typical spiral galaxies. Next, we generate a series of values for to obtain the graph of distribution speeds in the equatorial axis. We calculate these values at regular intervals from the origin, each one with 0.1 kpc, until to complete the length of 15 kpc. We have obtained the graph of this velocity distribution on a parallel plane to a distance of 0.2 kpc to the equatorial plane (see figure 4). In the equatorial plane ($\theta = 90^0$), the velocity distribution has a similar form to that of spherical symmetry distribution times a constant factor $[1-2\eta(2-\eta)/5]^{1/2}$ (see figure 1). We have adopted this zone for to obtain our graphic because on this plane is where the astronomical observations of their velocities are made.

A similar approaching is made for an oblate ellipsoidal distribution with radial densities.

For the next potential applying the QA we have:

$$\Phi = -\frac{GM}{3R^{5/2}}\left(5R^{3/2} - 2r^{3/2}\right) - \beta\frac{GM}{R^{5/2}}r^{3/2}\left(1 - 3Cos^2\theta\right), \tag{34}$$

and its corresponding gravitational field is:

$$g = -\frac{GMr^{1/2}}{R^{5/2}}\left\{\left[1 - \frac{3}{2}\beta(1 - 3Cos^2\theta)\right]\hat{r} - 6\beta Sin\theta Cos\theta\hat{\theta}\right\}, \tag{35}$$

and its distribution of matter writes ($n = 1/2$):

$$\rho = \left(\frac{M}{8\pi R^{5/2}}\right)\frac{1}{r^{1/2}}\left[5 + \frac{9}{2}\beta(1 - 3Cos^2\theta)\right]. \tag{36}$$

Here, $R$ is the value of the average radius. The $\beta$ value is defined for all the real numbers that fulfilled the condition that the amount locked up in the hook is positive. This determines the positive value of the equation of the density, because a negative value has not physical sense. It could be happened for the angles $\theta$ near or equal to zero. For this case, the interval of allowed $\beta$ values is $0 < \beta < 5/9$.

Now, as in the previous case, we can calculate the rotational velocity of stars with this radial distribution. Using the Newton's law, we have that

$$\frac{V^2}{rSin\theta} = \frac{GMSin\theta r^{1/2}}{R^{5/2}}\left[1 - \frac{1}{2}\beta(1 - 3Cos^2\theta) - 2\beta Cos^2\theta\right], \tag{37}$$

and the distribution of velocities result:

$$V(r,\theta)=\left(\frac{GM}{R}\right)^{1/2}\left(\frac{r}{R}\right)^{3/4} Sin\theta\left[1-\frac{1}{2}\beta(1+Cos^2\theta)\right]^{1/2}. \qquad (38)$$

From this equation we could to obtain its graphs for the rotational velocities of stars around the center of galaxy taking the case $\beta=0.1$. Also, we plotted this expression at intervals of $\Delta r=0.1$ kpc, until complete $r=15$ *kpc*.

Similarly to the previous case, we calculate the graph of velocities and densities distribution in one parallel plane to a distance of 0.2 kpc to the equatorial plane. The Figs. 5(a) and 6(a) represent the density and the velocities distribution, respectively. Similar to the previous case, the velocity distribution in this plane has a similar form to that of spherical symmetry distribution (at least until the radial distance $r=4.8$ *kpc*, see the curves 3(a) and 8(a)). This fact is reached because we have similar densities in these both cases, *i.e.*, the oblate ellipsoidal distribution of matter are also times a factor (5 + 9β/2) of that spherical symmetry (see the curves 7(a) and 2(a)). Same considerations will be following in the next cases for the different distribution of densities ($n=1, n=3/2, n=9/5$).

For the next potential:

$$\Phi=-\frac{GM}{R^2}(2R-r)-\beta\frac{GM}{R^2}r\left(1-3Cos^2\theta\right), \qquad (39)$$

In a similar form with the previous case we obtain its gravitational force:

$$g=-\frac{GM}{R^2}\left\{\left[1-\beta(1-3Cos^2\theta)\right]\hat{r}-6\beta Sin\theta Cos\theta\hat{\theta}\right\}, \qquad (40)$$

and the density distribution ($n=1$) is:

$$\rho=\left(\frac{M}{2\pi R^2}\right)\frac{1}{r}\left[1+2\beta(1-3Cos^2\theta)\right], \qquad (41)$$

where $0<\beta<1/4$. Using the Newton's law, its velocity distribution reads:

$$\frac{V^2}{rSin\theta} = \frac{GMSin\theta}{R^2}\left[1-\beta(1-3Cos^2\theta)-6\beta Cos^2\theta\right],$$

And therefore

$$V(r,\theta) = \left(\frac{GM}{R}\right)^{1/2}\left(\frac{r}{R}\right)^{1/2} Sin\theta\left[1-\beta(1+3Cos^2\theta)\right]^{1/2}. \qquad (42)$$

In the parallel plane at 0.2 kpc to the equatorial plane and $\beta = 0.1$, we have the distribution of matter (Eq. (41)), and velocities (Eq.(42)) according with the present results. They are showed by the curves 5(b) and 6(b) respectively. In the case of the equatorial plane, the distributions of densities and velocities are the curves 7(b) and 8(b) respectively.

In a same way, for the potential

$$\Phi = -\frac{GM}{R^{3/2}}\left(3R^{1/2} - 2r^{1/2}\right) - \beta\frac{GM}{R^{3/2}} r^{1/2}\left(1-3Cos^2\theta\right), \qquad (43)$$

their gravitational field, and its distribution of matter ( $n = 3/2$ ), are:

$$g = -\frac{GM}{R^{3/2}}\frac{1}{r^{1/2}}\left\{\left[1-\frac{\beta}{2}(1-3Cos^2\theta)\right]\hat{r} - 6\beta Sin\theta Cos\theta\hat{\theta}\right\}, \qquad (44)$$

and

$$\rho = \left(\frac{3M}{8\pi R^{3/2}}\right)\frac{1}{r^{3/2}}\left[1+\frac{7}{2}\beta(1-3Cos^2\theta)\right] \qquad (45)$$

respectively, with $0 < \beta < 1/7$. Using the Newton's law, the velocity distribution reads:

$$\frac{V^2}{rSin\theta} = \frac{GMSin\theta}{R^{3/2}}\frac{1}{r^{1/2}}\left[1-\frac{\beta}{2}(1-3Cos^2\theta)-6\beta Cos^2\theta\right],$$

and then

$$V(r,\theta)=\left(\frac{GM}{R}\right)^{1/2}\left(\frac{r}{R}\right)^{1/4} Sin\theta\left[1-\frac{\beta}{2}(1+9Cos^2\theta)\right]^{1/2}. \quad (46)$$

In comparison with the earlier results, we evaluated these quantities for the parallel plane at 0.2 kpc to the equatorial plane, and $\beta=0.1$. The graphs for their density (Eq. (45)) and its velocities (Eq. (46)) are presented with the curves 5(c) and 6(c) respectively. In the equatorial plane, the distribution of densities and velocities are the curve 7(c) and 8(c) respectively.

Finally, we have studied the potential:

$$\Phi=-\frac{GM}{R^{6/5}}\left(6R^{1/5}-5r^{1/5}\right)-\beta\frac{GM}{R^{6/5}}r^{1/5}\left(1-3Cos^2\theta\right), \quad (47)$$

with a gravitational field that we writes:

$$g=-\frac{GM}{R^{6/5}}\frac{1}{r^{4/5}}\left\{\left[1-\frac{\beta}{5}(1-3Cos^2\theta)\right]\hat{r}-6\beta Sin\theta Cos\theta\hat{\theta}\right\}, \quad (48)$$

and the corresponding density distribution ($n=9/5$) reads

$$\rho=\left(\frac{6M}{20\pi R}\right)\frac{1}{r^{9/5}}\left[1+\frac{24}{5}\beta(1-3Cos^2\theta)\right]. \quad (49)$$

We also observed that the allowed interval for $\beta$ is $0<\beta<0.4$. Calculating the velocity distribution for $\beta=0.1$, and the parallel plane at distance of *0.2 kpc* with respect to the equatorial plane we have:

$$V(r,\theta)=\left(\frac{GM}{R}\right)^{1/2}\left(\frac{r}{R}\right)^{1/10} Sin\theta\left[1-\frac{\beta}{5}(1-3Cos^2\theta)-6\beta Cos^2\theta\right]^{1/2}. \quad (50)$$

The curves 5(d) and 6(d) shown the density (Eq. (49)) and velocity distribution (Eq. (50)) respectively. For the equatorial plane, the distributions of densities and velocities are the curves 7(d) and 8(d) respectively.

## 4. Analysis and discussion

We have studied how the velocities of rotation depend of the symmetry of the matter distribution, and also of its density when this decreases inversely proportional to $n$th power of the distance from its centre (rotational axis). This function has a singularity, which goes to infinity when we approach to the origin (this could be meaning physically a supermassive black hole, how has been detected in many galaxies). With our analysis we obtain that the distributions of velocities grow very fast in this center as at the same time its speed decreases near the periphery, in agreement as its density grow near in its centre and decrease out of this. This velocity reaches a limit in the periphery when the density is larger (with $n \approx 1.8$). The form of this velocity curve is similar to that obtained by astronomical observations and, in addition, they are within their order of magnitude. The corresponding equation also shown as these rotational velocities are directly proportional to the root square of the mass and inversely proportional to the root square of equatorial radius. These properties are in direct relation with the results obtained by Rubin et al. [2]. They found that for galaxies of the same luminosity (proportional to the galaxy mass according with our deduction of the Tully Fisher law showed in Appendix A), the velocities are larger in more oval galaxies (Sa) than for galaxies more extended (Sb, Sc). Also for galaxies of the same type, but with different luminosity, their velocities decrease when their luminous intensity diminishes.

Also this result can help to explain why the stars near its centre of rotation (super massive black hole) turn very quickly, and those of the periphery are restrained to escape in spite of the high velocities reached. This great concentration of mass allows that the stars of the galaxies have a centre of gravitational rotation, avoiding that these be ejected outside of the galaxy.

## 5. Conclusion

In summary, the study of the dynamics of the galaxies conducts to rough estimate of the distribution of the rotational velocity of the stars with a similar form to those obtained by observation. Also we conclude that this method, besides doing unnecessary the hypothesis of the dark matter and modification of the Newton's laws, will serve to confirm the validity and reach of these laws to a greater spatial scale, and reinforce the weakest gravity premise. A better approximation will have to be carried out in the future with a more realistic approximation using computational simulation, but considering this type of analysis.

## Aknoledgements

E.L.S. is extremely grateful to Daniel Acosta Avalos and Arturo R. Samana for their very careful reading and helpful discussion of the manuscript.

E.L.S would like to acknowledge the partial support of Brazilian agencies: Conselho Nacional de Desenvolvimento Científico e Tecnológico (CNPq) and Centro Latino Americano de Física (CLAF).

## APPENDIX A: TULLY-FISHER LAW

The Tully-Fisher law (TFL), as we know, is an empirical relationship between the intrinsic luminosity (proportional to the luminous mass of a spiral galaxy) and its maximum velocity to the fourth power (Tully & Fisher, 1977). This discrepancy between luminosity and velocity was first detected by Zwicki (1937), when he found a difference between the solar masses of galaxies from Coma cluster derived of its light, and derived of gravity with the virial theorem using the velocity of these galaxies.

Until now the TFL doesn't have a mathematical deduction with first principles. We will analyze these galaxies with our model like a self-graviting system in a stationary state, so that the average kinetic and potential energy does not change with time, and therefore we can to use the virial theorem. We found with our model, that the better approximation to rotational curves of the stars in the galaxies is with a matter distribution $n\approx1.8$. We can to approximate its potential and kinetic energy for a star with mass $m$ rotating around a radius $R$, a distance of its center near to the periphery of the galaxy where it reaches its maximum velocity $V$:

$$K \propto mV^2 \quad \text{and} \quad U \propto -\frac{GmM}{R} \tag{A1}$$

therefore, according with the virial theorem

$$2\overline{K} = -\overline{U}\,, \tag{A2}$$

therefore we can write

$$V^2 \propto \frac{M}{R}\,. \tag{A3}$$

We know that the luminosity of a galaxy must to be proportional to its number of stars, i. e., its total mass

$$L \propto M\,. \tag{A4}$$

But its luminosity also depends on its superficial area, because according with a greater area the internal stars in the galaxy are less covered, and they have greater possibilities of than its light leaves out of its surface, and increases its luminosity. Therefore

$$L \propto R^2\,. \tag{A5}$$

Then, we can replace these relations, (A4) and (A5), in the relation (A3) for to obtain the TFL:

$$L \propto V^4\,. \tag{A6}$$

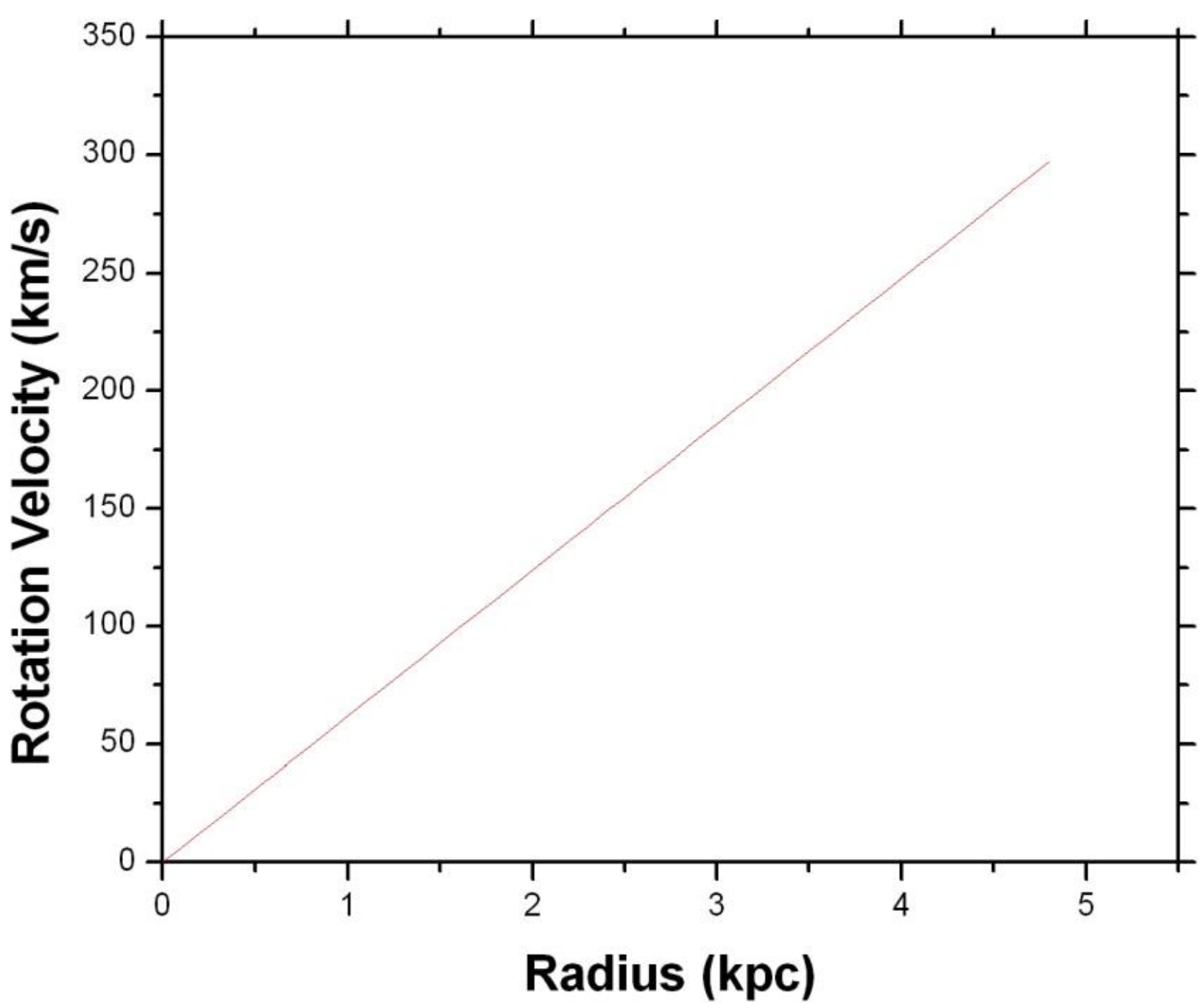

Figure 1. Rotational velocity of a spherical distribution of matter with constant density.

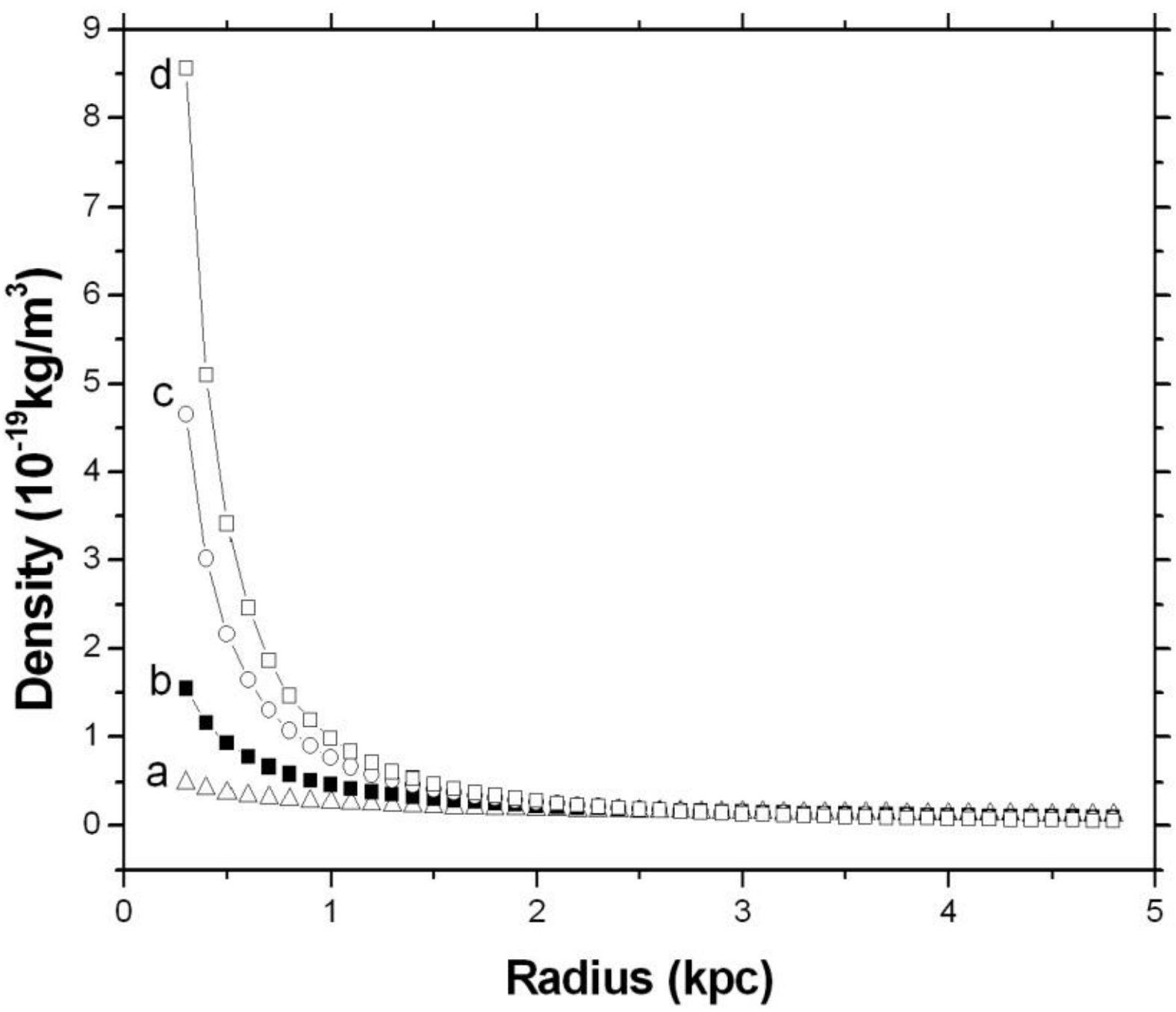

Figure 2. Density of a spherical distribution of matter that grows how the inverse of the radius elevated to some *n*-power: (a) $n=1/2$, (b) $n=1$, (c) $n=3/2$, and (d) $n=9/5$.

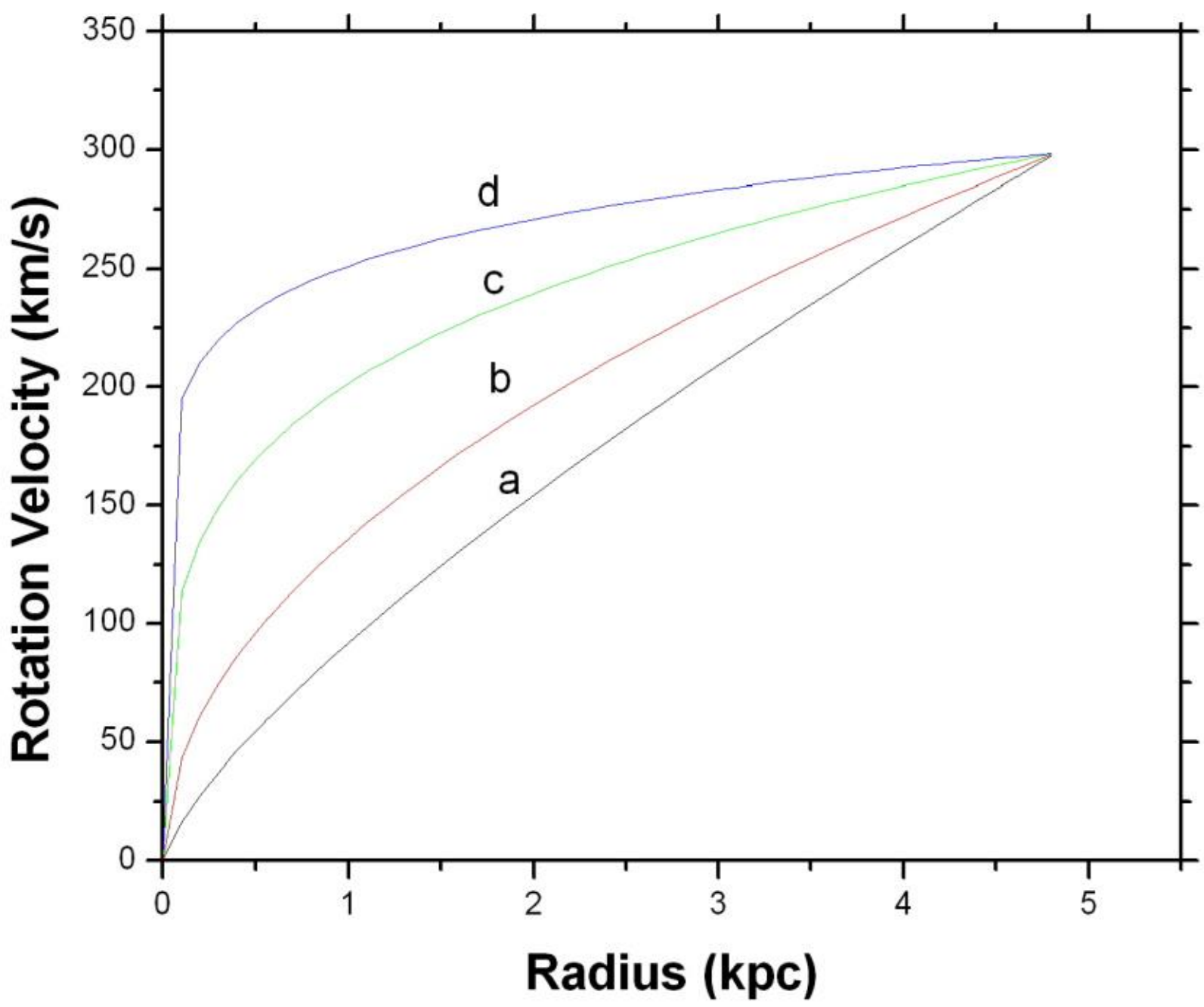

Figure 3. Rotational velocities of a spherical distribution of matter with the density of r inversely proportional to some *n*-power: (a) $n = 1/2$, (b) $n = 1$, (c) $n = 3/2$, and (d) $n = 9/5$, in the same points where we evaluated the density shown in figure 2.

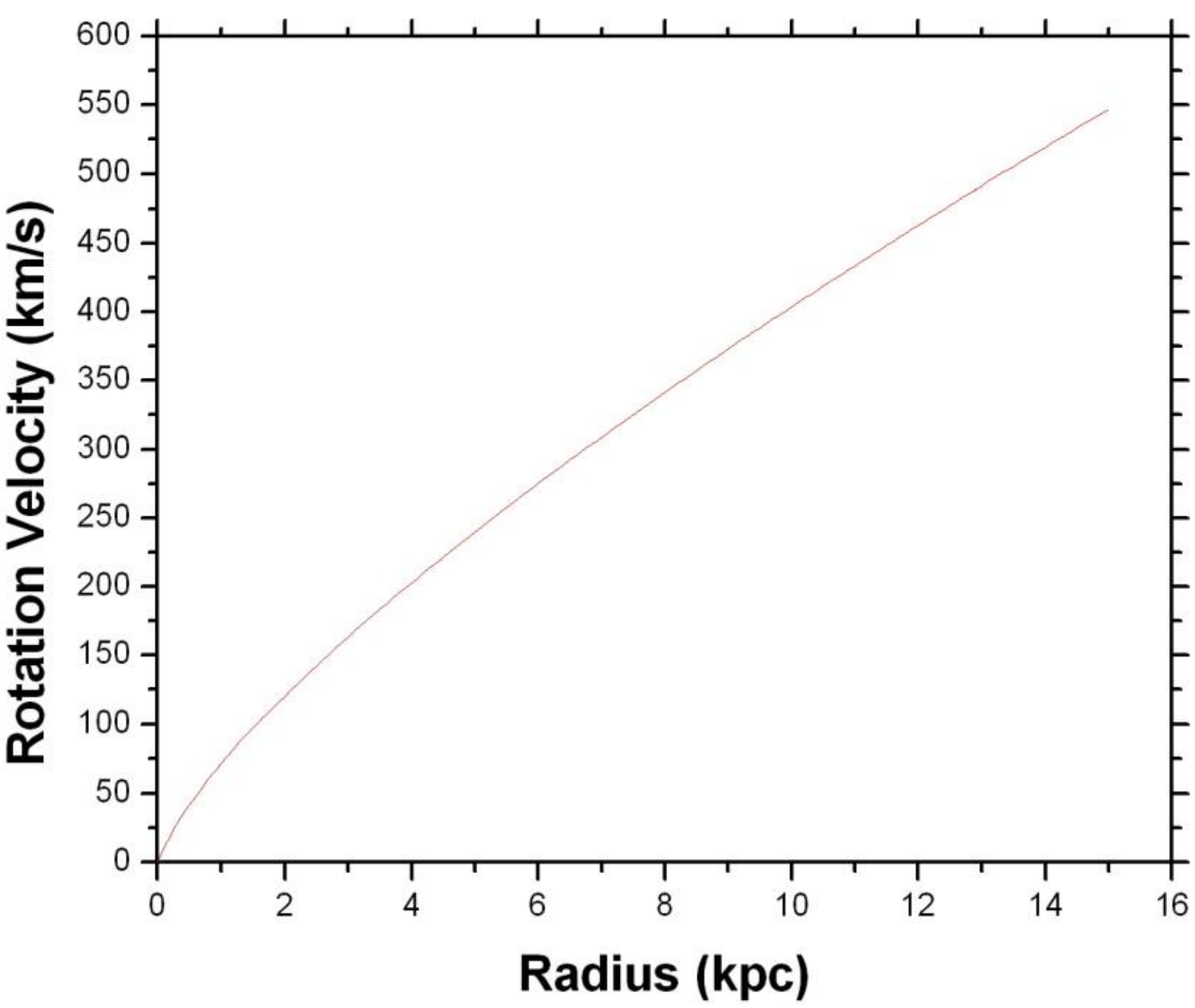

Figure 4. Rotational velocities of an oblate ellipsoid distribution of matter with constant density.

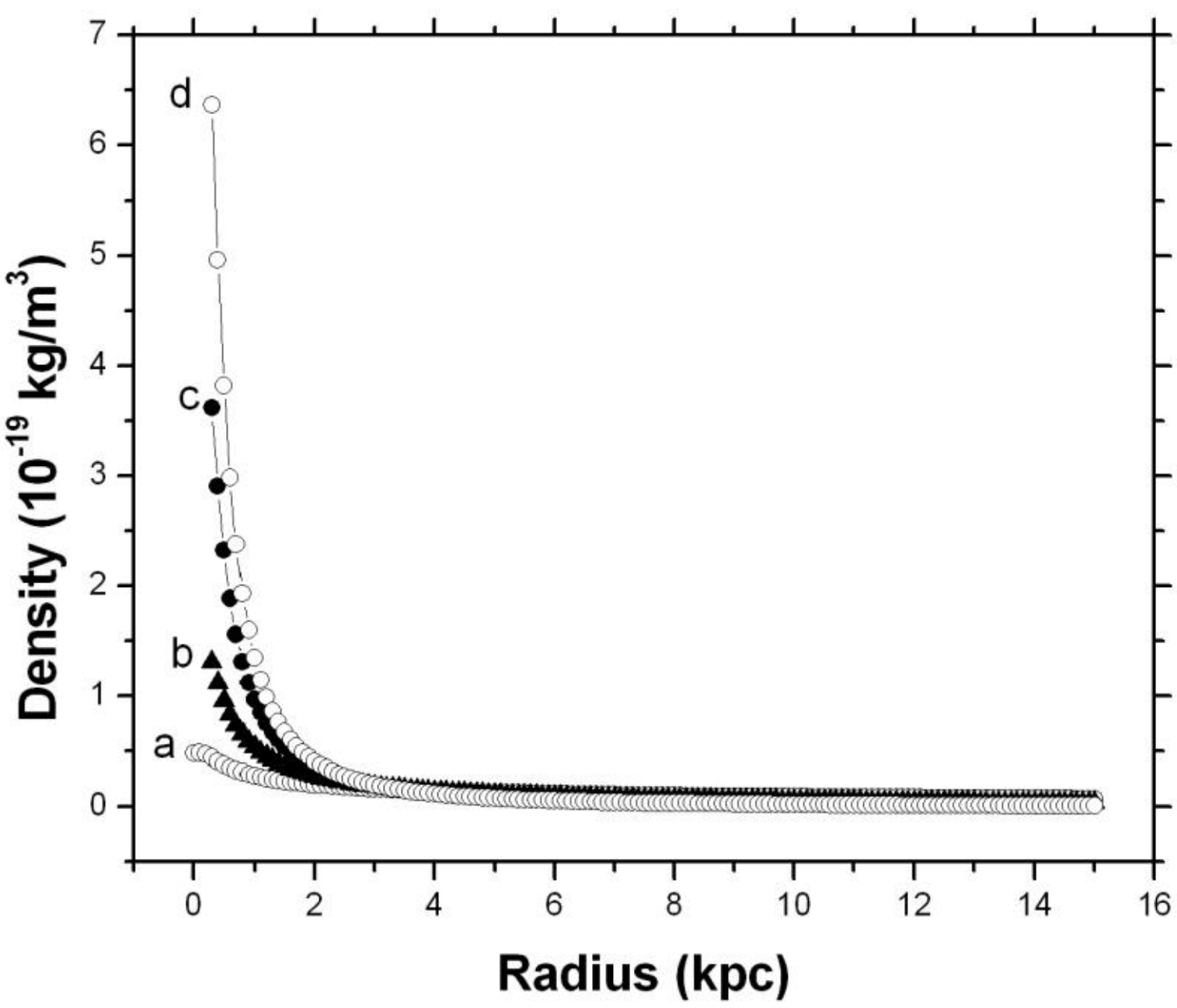

Figure 5. Densities of an oblate ellipsoidal distribution of matter that grows as function of $\theta$, and $r$ inversely proportional to some $n$-power: (a) $n = 1/2$, (b) $n = 1$, (c) $n = 3/2$, and (d) $n = 9/5$.

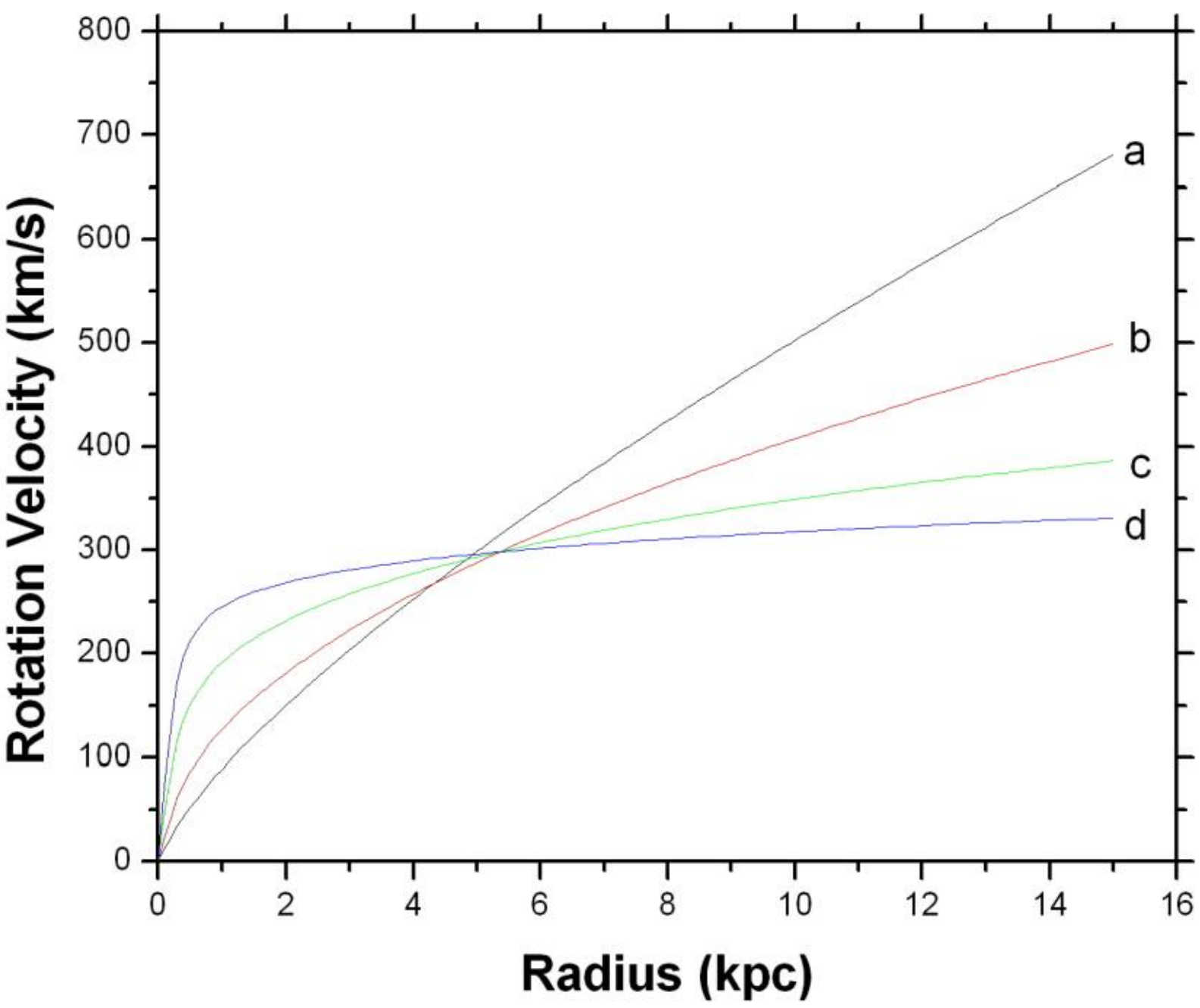

Figure 6. Rotational velocities of an oblate ellipsoid distribution of matter as function of $\theta$, and $r$ inversely proportional to some $n$th-power: (a) $n = 1/2$, (b) $n = 1$, (c) $n = 3/2$, and (d) $n = 9/5$, in the same points where we evaluated the density shown in figure 5.

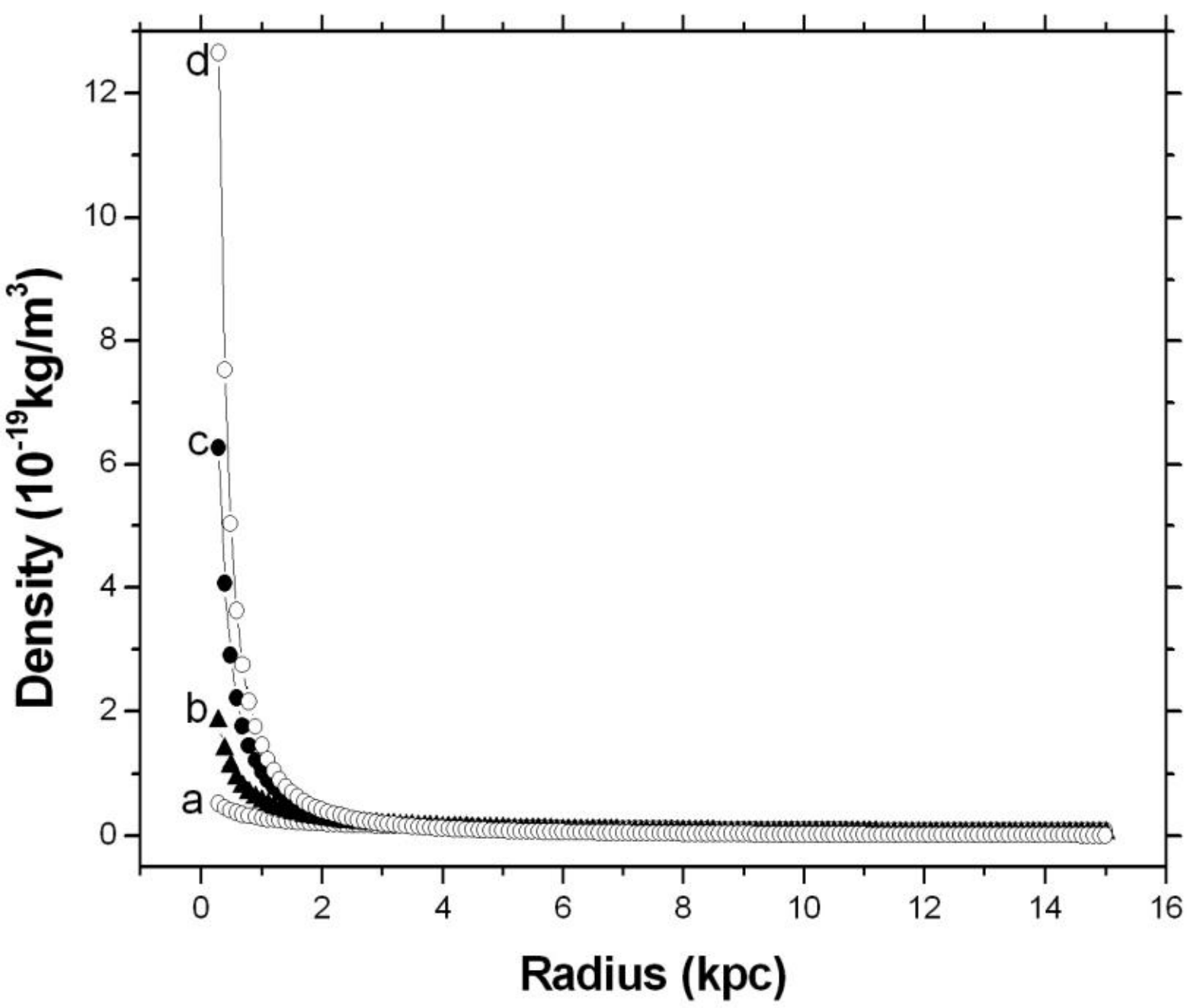

Figure 7. Density on the equatorial plane ($\theta = 90^0$) of an oblate ellipsoid distribution of matter as a function of $r$ to some $n$th-power: (a) $n = 1/2$, (b) $n = 1$, (c) $n = 3/2$, and (d) $n = 9/5$.

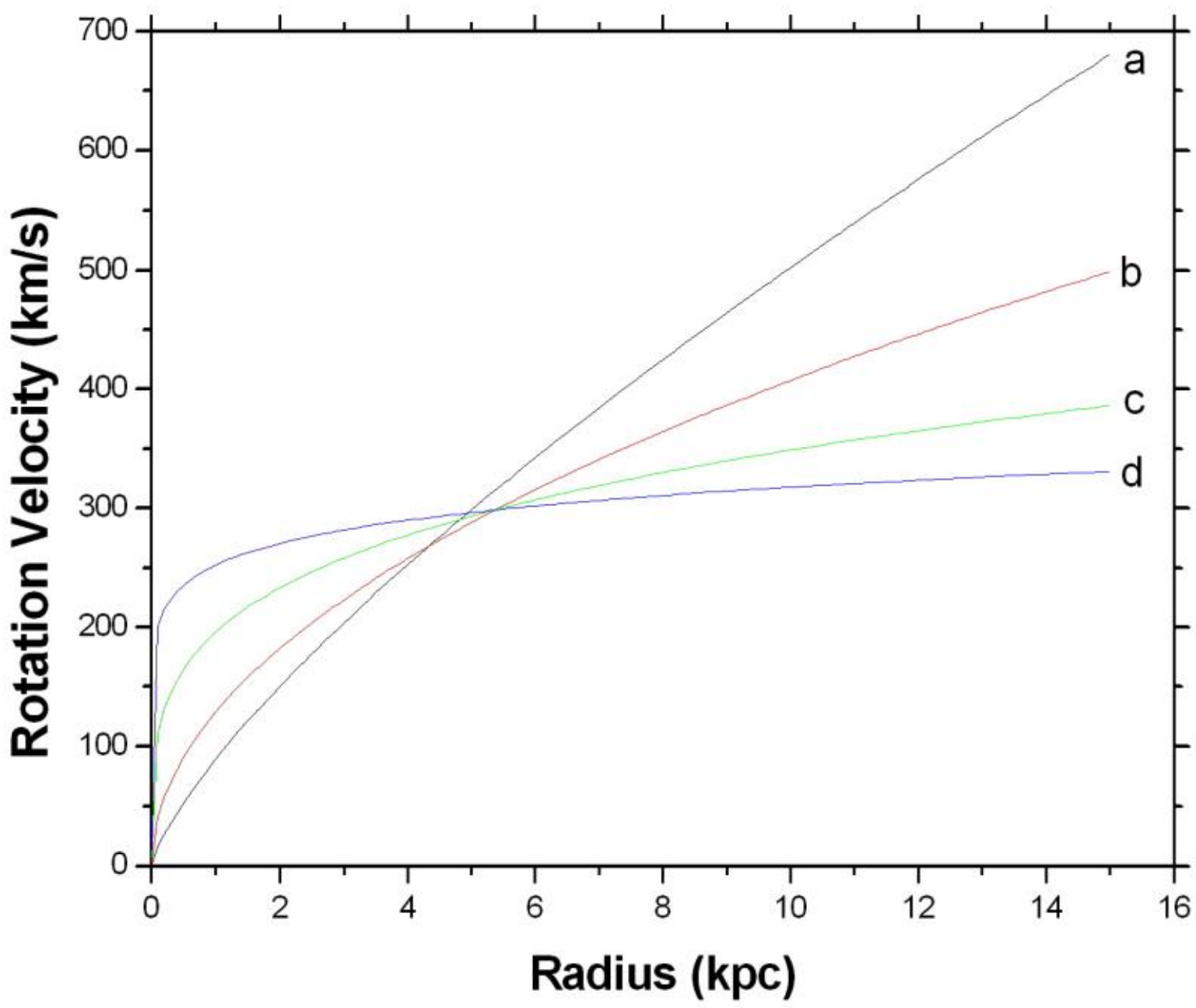

Figure 8. Rotational velocities on the equatorial plane ($\theta = 90^0$) of an oblate ellipsoid distribution of matter as function of $r$ inversely proportional to some $n$th-power: (a) $n = 1/2$, (b) $n = 1$, (c) $n = 3/2$, and (d) $n = 9/5$, in the same points where we evaluated the density shown in figure 7.